\documentstyle [12pt] {article}
\title{SINGLE CAPACITOR PARADOX}

\author{Vladan Pankovi\'c\\
Department of Physics, Faculty of Sciences, 21000 Novi Sad,\\ Trg
Dositeja Obradovi\'ca 4. , Serbia, vpankovic@if.ns.ac.yu}

\date {}
\begin{document}
\maketitle \vspace {0.5cm}
 \vspace {0.5cm}

\begin {abstract}
In this work single capacitor paradox (a variation of the
remarkable two capacitor paradox) is considered. Simply speaking
in an ideal (without any electrical resistance and inductivity)
electrical circuit with single charged capacitor and switch, by
transition from initial, open state (switch in OFF position) in
the final, closed state (switch in ON position), there is a total
loss of the initial energy of the electrical field in condenser.
Given energy loss can be simply explained without any dissipative
effects (Joule heating or electromagnetic waves emission) by work
of the electrical field by movement of the charge from one in the
other plate of the capacitor. (Two capacitors paradox can be,
obviously, explained in the analogous way.)
\end {abstract}

\vspace {1.5cm}

As it is well-known remarkable two-capacitors paradox, formulated
and considered in many textbooks and articles on the basic
principles and applications of the electronic and electrodynamics
[1]-[7], states the following. Consider an ideal (without any
electrical resistance and inductivity) electrical circuit with
first, initially charged, and second, initially non-charged, of
two identical capacitors. In given circuit, by transition from
initial, open state (switch OFF state) in the closed state (switch
ON state), an unexpected, mysterious loss of the half of initial
energy of electrical fields within capacitors occurs. Different
authors [4]-[7] suggest that given energy loss is realized by
different dissipative processes (Joule heating or/and
electromagnetic waves emissions) realized by non-neglectable
residual electric resistances and inductivities in realistic
circuits.

In this work single capacitor paradox (a variation of the two
capacitor paradox) will be considered. Simply speaking in an ideal
(without any electrical resistance and inductivity) electrical
circuit with single charged capacitor and switch, by transition
from initial, open state (switch OFF position) in the final,
closed state (switch ON position), there is a total loss of the
initial energy of the electrical field in the condenser. Given
energy loss can be simply explained without any dissipative
effects (Joule heating or electromagnetic waves emission) by work
of the electrical field by movement of the charge from one in the
other plate of the capacitor. (Two capacitors paradox can be,
obviously, explained in the analogous way, which goes over basic
intention of this work.)

Consider a simple an ideal (without any electrical resistance and
inductivity) electrical circuit with single charged capacitor with
the capacitance C and one switch.

Initially, switch is in the state OFF so that electrical circuit
is open. Then capacitor is charged by electrical charge Q and
holds voltage $V=\frac {Q}{C}$. Energy of the electric field
within capacitor equals, as it is well-known,
\begin {equation}
 E_{in}= \frac { CV^{2}}{2} = \frac {Q^{2}}{2C}
\end {equation}

But when switch turns out in the state ON electrical circuit
becomes closed and during a very small time interval capacitor
becomes uncharged. Then, energy of the electric field within
capacitor becomes zero, i.e.
\begin {equation}
 E_{fin}= 0         .
\end {equation}

In this way there is the following energy difference between the
initial and final state of  given electrical circuit
\begin {equation}
 \Delta E = E_{fin} - E_{in}= - E_{in} = - \frac {Q^{2}}{2C} .
\end {equation}
It seems as a paradoxical total energy loss.

Consider now total energy of the electrical field of capacitor
more accurately.

When capacitor holds some charge q, for $0 \leq q \leq Q$, and
corresponding voltage $v=\frac {q}{C}$,   energy of the electrical
field within capacitor equals
\begin {equation}
  E= \frac {q^{2}}{2C}    .
\end {equation}

Suppose now that, by action of the electrical field, charge of the
capacitor decreases for infinite small value dq that turns out
from the one in the other capacitor plate. Then total energy of
the electrical field of capacitor becomes
\begin {equation}
  E + dE = \frac {(q-dq)^{2}}{2C} =  \frac {q^{2}}{2C}- \frac {q dq}{C}=  \frac {q^{2}}{2C}- q dv .
\end {equation}
where small term proportional to $(\delta q)^{2}$ is neglected. As
it is not hard to see term $q \frac {dq}{C}$ in (5) can be
presented in the form (dq)v that, according to well-known
definitionof the work in the electric field, can be considered as
the work done by the electrical field by movement of the
electrical charge dq from one in the other capacitor plate.

According to (4), (5) it follows
\begin {equation}
  dE =  -q dv = - \frac {q dq}{C}
\end {equation}
that, obviously, represents law of the conservation of electrical
field energy. In other words, diminishing of the electrical field
energy is equivalent to work done by  electrical field by
movement of the electrical charge from one in the other capacitor
plate.

After simple integration of (6) over [0, Q] interval of q values,
it follows
\begin {equation}
   \Delta E = - \frac {Q^{2}}{2C}       .
\end {equation}
It is, obviously, identical to (3).

In this way we obtain very simple and reasonable solution of given
single capacitor paradox in the completely ideal (without any
electrical resistance or inductivity) electrical circuit. (More
precisely we shall consider that really existing resistance and
inductivities yield only high order corrections which here can be
neglected.)

In conclusion, the following can be shortly repeated and pointed
out. In this work we suggest simple solution of the single
capacitor paradox in the completely ideal (without any electrical
resistance or inductivity) electrical circuit.  Namely, it is
shown that total energy loss of the electrical field within
capacitor corresponds to well-known work done by electrical field
by movement of the electrical charge from one in the other
capacitor plate in full agreement with energy conservation law.
Nothing more (e.g. some dissipative processes, Joule heating and
electromagnetic wave emission effects) is necessary. (Two
capacitors paradox can be, obviously, explained in the analogous
way, which goes over basic intention of this work.)

\vspace{1cm}

Author is deeply grateful to Prof. Dr. Darko Kapor for
illuminating discussions.

\vspace{1cm}

 {\large \bf References}

\begin {itemize}

\item [[1]] D. Halliday, R. Resnick, {\it Physics, Vol. II} (J. Willey, New York, 1978)
\item [[2]] F. W. Sears, M.W. Zemansky, {\it University Physics} (Addison-Wesley, Reading, MA, 1964)
\item [[3]] M. A. Plonus, {\it Applied Electromagnetics}, (McGraw-Hill, New York, 1978)
\item [[4]] E. M. Purcell, {\it Electricity and Magnetism, Berkeley Physics Course Vol. II} (McGraw-Hill, New York, 1965)
\item [[5]] R. A. Powel, {\it Two-capacitor problem: A more realistic view}, Am. J. Phys. {\bf 47} (1979) 460
\item [[6]] T. B. Boykin, D. Hite, N. Singh, Am. J. Phys. {\bf 70} (2002) 460
\item [[7]] K. T. McDonald, {\it A Capacitor Paradox}, class-ph/0312031

\end {itemize}

\end {document}